\documentclass[chi_draft]{sigchi}

\toappear{
Permission to make digital or hard copies of all or part of this work
for personal or classroom use is granted without fee provided that
copies are not made or distributed for profit or commercial advantage
and that copies bear this notice and the full citation on the first
page. Copyrights for components of this work owned by others than the author(s)
must be honored. Abstracting with credit is permitted. To copy
otherwise, or republish, to post on servers or to redistribute to
lists, requires prior specific permission and/or a fee. Request
permissions from \href{mailto:jyzhou15@cs.washington.edu}{jyzhou15@cs.washington.edu}. \\
ACM xxx-x-xxxx-xxxx-x/xx/xx\ldots \$15.00 \\
DOI: \url{https://dx.doi.org/xx.xxxx/xxxxxxx.xxxxxxx}
}

\usepackage{balance}       
\usepackage{graphics}      
\usepackage[T1]{fontenc}   
\usepackage{txfonts}
\usepackage{mathptmx}
\usepackage[pdflang={en-US},pdftex]{hyperref}
\usepackage{color}
\usepackage{booktabs}
\usepackage{textcomp}

\usepackage{microtype}        
\usepackage{ccicons}          

\usepackage{csquotes}         
\usepackage{forest}           
\usepackage{enumitem}         
\usepackage{amsmath}          
\usepackage{amssymb}          
\usepackage{dblfloatfix}      

\usepackage{todonotes}

\def\plaintitle{An Interactive UI to Support Sensemaking over Collections of Parallel Texts}
\def\plainauthor{Joyce Zhou, Elena Glassman, Daniel S. Weld}

\def\plainkeywords{
    mixed-initative system;
    sensemaking;
    text alignment;
    multiple alignment.
}

\makeatletter
\def\url@leostyle{%
  \@ifundefined{selectfont}{
    \def\UrlFont{\sf}
  }{
    \def\UrlFont{\small\bf\ttfamily}
  }}
\makeatother
\def\pprw{8.5in}
\def\pprh{11in}

\definecolor{linkColor}{RGB}{6,125,233}
\hypersetup{%
  pdftitle={\plaintitle},
  pdfauthor={\plainauthor},
  pdfkeywords={\plainkeywords},
  pdfdisplaydoctitle=true, 
  bookmarksnumbered,
  pdfstartview={FitH},
  colorlinks,
  citecolor=black,
  filecolor=black,
  linkcolor=black,
  urlcolor=linkColor,
  breaklinks=true,
  hypertexnames=false
}

\newcommand{\sys}{\mbox{\sc Avtaler}} 

\begin{document}

\title{\plaintitle}


\numberofauthors{3}
\author{
  \alignauthor{Joyce Zhou\\
    \email{jyzhou15@cs.washington.edu}}\\
    \affaddr{University of Washington}\\
  \alignauthor{Elena Glassman\\ 
    \email{glassman@seas.harvard.edu}}\\
    \affaddr{Harvard University}\\
  \alignauthor{Daniel S. Weld\\
    \email{weld@cs.washington.edu}}\\
    \affaddr{University of Washington}\\
}

\maketitle

\begin{abstract}
  Scientists and science journalists, among others, often need to make sense of a large number of papers and how they compare with each other in scope, focus, findings, or any other important factors.
  However, with a large corpus of papers, it’s cognitively demanding to pairwise compare and contrast them all with each other.
  Fully automating this review process would be infeasible, because it often requires domain-specific knowledge, as well as understanding what the context and motivations for the review are.
  While there are existing tools to help with the process of organizing and annotating papers for literature reviews, at the core they still rely on people to serially read through papers and manually make sense of relevant information.

  We present \sys\footnote{Norwegian for "to make an agreement on"}, which combines peoples' unique skills, contextual awareness, and knowledge, together with the strength of automation.
  Given a set of comparable text excerpts from a paper corpus, it supports users in sensemaking and contrasting paper attributes by interactively aligning text excerpts in a table so that comparable details are presented in a shared column.
  \sys\ is based on a core alignment algorithm that makes use of modern NLP tools.
  Furthermore, \sys\ is a mixed-initiative system: users can interactively give the system constraints which are integrated into the alignment construction process.
\end{abstract}






\section{Introduction}
\label{sec:introduction}

\begin{figure*}[htb!]
    \label{fig:overview}
    \centering
    \includegraphics[width=1.75\columnwidth]{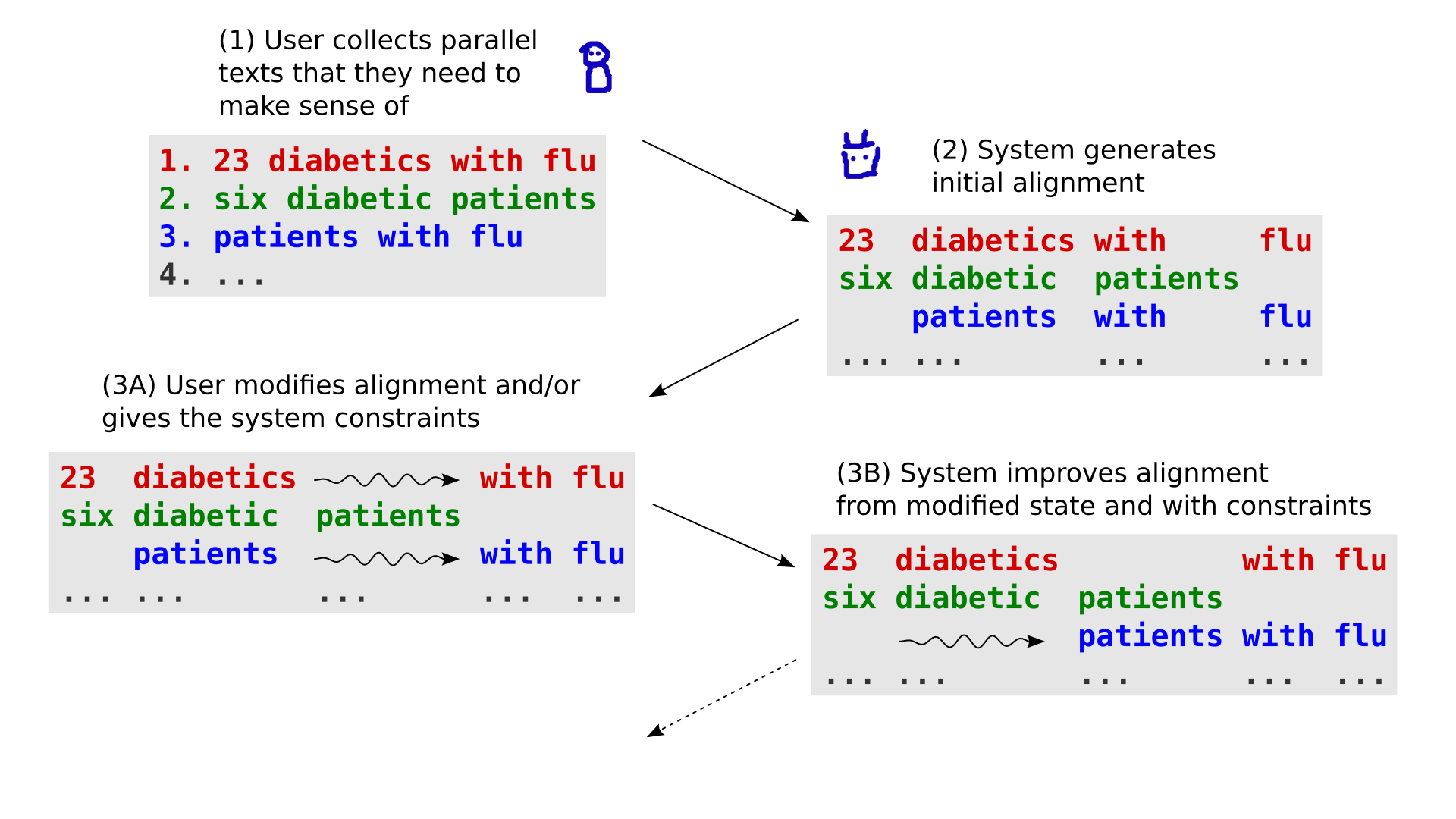}
    \caption{
      An overview of the mixed-initiative system structure.
      An more detailed example of potential user-system interaction cycle and motivations in section "\nameref{sec:usagescenario}".
      Detail about system implementation in section "\nameref{sec:systemdesign}".
      Detail about system interface and user controls in section "\nameref{sec:systemdesign}".
    }
\end{figure*}

Scientists and science journalists, among others, need to understand what has been investigated and concluded in past work.
For example, a journalist may want to summarize how numerous different papers describe the social effects of a historical event as part of their background reading.
A student might want to make sense of what ways different machine learning algorithms have been modified, improved, or applied.
A medical researcher could review previous work on a specific medical condition because they want to know which populations or coexisting conditions have been studied.
In general, this reviewing and understanding means reading through and comparing vast amounts of different types of information, for example: methods, questions, datasets, results.
However, especially in fast-moving areas of scientific inquiry, it is time-consuming and cognitively demanding to read and remember the details necessary to build a mental model of the relationships between all the relevant papers.

Current tools for managing large literature reviews, such as Zotero, help readers by indexing papers using metadata and user-generated annotations, but they still rely on the readers to serially read through papers and manually make sense of them.
There has also been work on identifying and extracting segments of paper abstracts based on sentence purpose, such as the PICO\_Parser for the PICO (Patient-Intervention-Comparison-Outcome) medical paper abstract structure, or building datasets like CODA-19.
However, while these tools help with extracting and preserving relevant paper details, readers still need to make a significant amount of effort to make sense of the extracted information.

This brings us to the main motivations behind this project.
How do we support users who need to build a schema to describe different aspects of a huge collection of text extracted from scientific papers, from scratch?
What kind of integrated AI assistance would be most helpful for making sense of extracted information?

For example, imagine a researcher has collected a large collection of papers on breast cancer treatments and they want to make sense of them as a collection while hoping to avoid reading thoroughly through all of them.
Making sense of them as a collection means being able to answer questions like, \textit{Who was represented in these studies? What treatments were tried? What outcomes were tracked, and what were their associated values?}
They may have already extracted phrases from the paper abstracts using a system like PICO\_Parser, saving the relevant ones as a list of parallel text phrases that are all individually tied to the question they are interested in.
However, it is still daunting to read the long list of parallel texts and manually or mentally compiling a distribution that answers the questions above.

Our approach is building a mixed-initiative system for aligning these parallel texts into a table where comparable details are presented in a shared column, to support comprehension of cross-document similarities and differences.

We wrote an algorithm that can build alignments out of a list of parallel texts where each row in the alignment represents a single text ("multiple alignments").
To complement this algorithm, we wrote a stochastic search algorithm that can improve an existing multiple alignment, centered around original heuristic alignment quality score.

But because these parallel scientific texts are in a specialized domain, a fully-automated system may not capture all of the details that a researcher recognizes and cares about.
Therefore, we keep the human in the loop, giving them controls that can change the multiple alignment and system search directly.
Thus, \sys\ provides automated support while still allowing the human users to make changes and give feedback to the system, to get the most benefit from each side.

\section{Background and Related Work}
\label{sec:background}

There has been a huge amount of work on information overload.
Some researchers have worked on tools for quickly understanding single papers in isolation \cite{Cachola2020TLDRES}, but we focus on tools for making sense of \emph{multiple} papers in context and building an understanding of how they relate to each other.

The task of understanding sets of papers can be broken down into building collections of related papers (which is often helped along by library assistants such as Zotero and Mendeley, or search engines such as Semantic Scholar or Google Scholar) and then sensemaking of the papers themselves.
We focus on sensemaking of a pre-collected group of papers in context.

\subsection{Parallel Text Structure and Extraction}
\label{sec:background_parallelstructure}

In order to make sense of a set of related papers, people typically start with some kind of high-level structure or schema to organize paper information with.
These structures vary depending on the context or content they are designed for.
For example, CODA-19 \cite{Huang2020CODA19RA} codes each sentence of the CORD-19 paper database abstracts into Background, Purpose, Method, Findings, and "Other" categories.
The PICO search strategy tool \cite{Eriksen2018TheIO} is designed specifically for medical papers and segments abstracts into Population, Intervention, Comparison, and Outcome.
Alternative tools such as SPIDER \cite{Cooke2012BeyondP} and PICOS \cite{Methley2014PICOPA} have also been proposed to better handle qualitative and mixed-methods research papers.

Of course, not all papers necessarily have this information clearly delineated or identified.\footnote{Building the CODA-19 dataset was a manual, crowd-sourced effort!}
Even if they are clearly identified, there are frequently variations in the exact schema being used.
Thus, there is previous work that focuses on automatically identifying and extracting relevant segments of paper abstracts and contents.
For example, PICO\_Parser \cite{Kang2019PretrainingTR} uses deep NLP methods to identify PICO elements of paper abstracts and standardize references to medical concepts using the UMLS classification system.
"Supervised Distant Supervision" is a similar approach to the same task, focusing on supplementing the training data to improve the ML model \cite{Wallace2016ExtractingPS}.

A different common approach to structuring paper information for sensemaking is based on using paper metadata in addition to (or in place of) the paper abstract or contents themselves.
This includes but is not limited to extracting or predicting citation networks \cite{Yasunaga2019ScisummNetAL, Hope2020SciSightCF, Hosseini2018AnalysisOC}, citation sentences \cite{Elkiss2008BlindMA, Zhou2020MultilevelTA}, co-authorship \cite{Hope2020SciSightCF, Jin2021CommunityDA}, or co-mentions of specific topics \cite{Hope2020SciSightCF}.

Finally, parallel structure extraction is not a task unique to paper reading.
There has been plenty of related work in other contexts, such as software engineering and linguistics.
For instance, Examplore \cite{Glassman2018VisualizingAU} builds a code skeleton with API usage features around a given API method in order to visualize how to use the method.
METER ("MEasuring TExt Reuse") \cite{Clough2002METERMT} is a project that, while not fully extracting structure, uses n-gram overlap, greedy string tiling, and a cognate-based alignment algorithm \cite{Simard1993UsingCT} to evaluate text reuse in news agency copy text.

\subsection{Visualization}
\label{sec:background_visualization}

Building good visualizations of literature review information is also very important, as the ultimate goal is for the reader to be able to make sense of it.
Existing visualizations range widely in the data they rely on, the purpose they are designed for, and the style of the visualizations themselves.
For example, one system visualizes a set of PICO "Interventions" covered by a given set of papers with a hierarchical tree where similar intervention styles are grouped together \cite{Vong2014VisualizationOP}.
SciSight\footnote{\href{https://scisight.apps.allenai.org/}{https://scisight.apps.allenai.org/}}, which is designed more as an exploratory tool, visualizes the frequency that extracted medical terms are collocated, as well as paper co-authorship with a graph where author groups are represented with nodes \cite{Hope2020SciSightCF}.
WordSeer\footnote{\href{https://wordseer.berkeley.edu/}{https://wordseer.berkeley.edu/}}, which is intended for use as an exploratory tool in digital humanities, graphs how frequently specific search terms are used in specific contexts (e.g. allowing users to see how often different presidents were referred to with an adjective synonymous with "great") \cite{Muralidharan2013WordSeerAK}.

However, literature review is not the only field where visualizing large sets of parallel texts is valuable.
In philology, the study of historical development of literary texts, one of the most common tasks is to highlight correspondences and divergences between different versions of texts originating from the same sources.
For instance, comparisons of different versions of the same Bible verse are fairly common.
Visualizations for text collation usually use some kind of text alignment in a table or "subway map" format which emphasize which segments of texts are identical or different.
For example, iAligner\footnote{\href{https://ialigner.com/}{https://ialigner.com/}} presents each input text in its own row with identical substrings maximally aligned together and highlights segments with nonzero edit distances \cite{Yousef2016iAlignerAT}.
TraviZ\footnote{\href{https://www.traviz.vizcovery.org/}{https://www.traviz.vizcovery.org/}} presents an interactive "subway map" style visualization with each shared or distinct substring roughly in order of where they appear, and lines connecting each substring to represent each original input text \cite{Jnicke2015TRAVizAV}.
CollateX\footnote{\href{https://collatex.net/}{https://collatex.net/}} actually outputs a lower-level general alignment format that is intended to be adapted to different visualization styles \cite{Dekker2015ComputersupportedCO}.
Similar table alignment styles have been used for phonetic comparisons \cite{Prokic2009MultipleSA}, machine translation \cite{MacCartney2008APA}, and rendering "word sequence variation patterns" \cite{Meng2011DeterminingWS}.

\section{Usage Scenario}
\label{sec:usagescenario}

We present an example usage scenario for \sys\ to illustrate why a user might choose to use this system, how they would use it, and how the user interface facilitates different features that a user may want in the process of sensemaking parallel texts.

Note that for ease of readability in this paper, we present images of alignments in a table spreadsheet format in this section.
These are not screenshots of \sys\ itself.
However, these are representative of results that \sys\ produces.
Descriptions of the actual \sys\ interface are described in the "User Interface" section.

\subsection{Problem Setting / Motivation}
\label{sec:usagescenario_motivation}

Kim is a researcher interested in the intersection between ASD and anxiety.
They are doing a literature review and have assembled a list of existing papers potentially studying these conditions.
Now they want to make sense of which participant populations have been studied in these past research papers in order to be able to summarize it to others and gain a better idea of what gaps in understanding there might be.

\subsection{Collecting Parallel Texts}
\label{sec:usagescenario_collectdata}

\begin{figure*}[htb!]
    \centering
    \includegraphics[width=2\columnwidth]{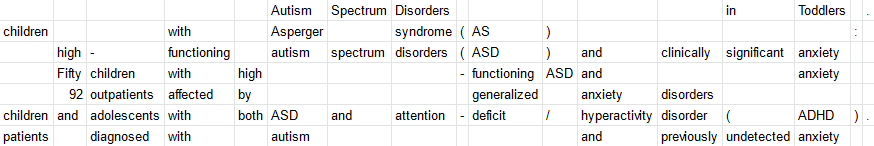}
    \caption{
        Usage Scenario: the full initial alignment of the input texts generated by \sys.
        This starting state will be improved by a mixture of human suggestions and autonomous optimizations.
    }
    \label{fig:usage_demo_1_3}
\end{figure*}

Kim starts by reviewing the parallel text schema that are typically used by authors of medical papers and decides that the "Participant" category of the PICO schema is most relevant to their goals.
While most papers have sections in their abstracts explicitly subtitled "Participant" that describes the populations studied, these sections vary a lot in length and structure.
Furthermore, some papers use a different schema entirely: they may not have this section explicitly separated out at all, or split important details about participant populations into multiple fine-grained sections.

Kim uses a tool (e.g. PICO\_Parser) to automatically extract phrases from each abstract describing the participant populations.
Now Kim has a dataset containing some number of plaintext phrases from each paper they're reviewing that describes the populations that each study focuses on.
Many of these phrases still vary in structure and level of detail, so Kim wants to get a sense of which types of detail are described frequently and, ideally, a breakdown of how different subgroups of each detail type are studied.

This is a sample of the texts that Kim has collected:

\begin{enumerate}[nosep]
    \item Autism Spectrum Disorders in Toddlers .
    \item children with Asperger syndrome ( AS ) :
    \item high - functioning autism spectrum disorders ( ASD ) and clinically significant anxiety
    \item Fifty children with high - functioning ASD and anxiety
    \item 92 outpatients affected by generalized anxiety disorders
    \item children and adolescents with both ASD and attention - deficit / hyperactivity disorder ( ADHD ) .
    \item patients diagnosed with autism and previously undetected anxiety
\end{enumerate}

Note that it is possible to work with a larger collection of text phrases.
However, to make this usage scenario easy to read, we will show alignments containing only this sample.

\subsection{Initial Alignment}
\label{sec:usagescenario_initial}

Kim decides to use \sys\ to visualize this participant dataset.
They paste the list of plaintext phrases they want to visualize into the data entry box and click "Align", then waits for \sys\ to finish generating a starting alignment.


The starting alignment is viewable in Figure \ref{fig:usage_demo_1_3}.

\subsection{Interactive Sensemaking}
\label{sec:usagescenario_interactive}

Once \sys\ finishes generating the starting alignment, it starts an interactive session where it renders the alignment and allows Kim to manually edit the alignment, change alignment search constraints, and tell the system to revise the alignment based on their changes.
As Kim is looking at the starting alignment, they may want to make changes if they notice alignment cells with content that should be combined together, single columns with terms that shouldn't be aligned, multiple columns that describe comparable concepts, or other types of flaws in the alignment.
They may want to modify the alignment search constraints because they are certain about an content change they made and want to prevent any chance of \sys\ reverting it. For instance, they may lock a column in an alignment to prevent \sys\ from modifying it or re-ordering any other content in relation to it.

\subsubsection{Merging text}
\label{sec:usagescenario_interactive_mergecell}

Kim notices that in the starting alignment, some parts of the text were tokenized in ways that end up separating texts that consist of single acronyms inside parentheses (such as "(ASD)" or "(ADHD)") into multiple cells.
In the context that Kim is working in, they expect these acronyms with or without parentheses to be functionally equivalent.
So, for each group of cells that make up a term, Kim merges the cell contents into a single cell.

Kim also notices that one of the input texts references a condition with a long name, which is also split into multiple cells.
Again, because Kim knows that the condition name is distinct and wouldn't make sense to keep separated across multiple alignable cells, they decide to merge the entire condition name together into a single cell.

\begin{center}
    \label{fig:usage_demo_2_1}
    \includegraphics[width=\columnwidth]{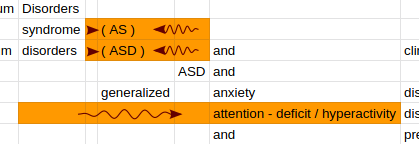}
\end{center}

\subsubsection{Deleting columns}
\label{sec:usagescenario_interactive_deletecol}

Kim also notices that some of the input texts have extra punctuation at the end of the lines, which aren't necessary to keep in the alignment (and may actually worsen readability if handled poorly by accident).
Kim chooses to delete the columns that contain these extra punctuation cells.

\begin{center}
    \label{fig:usage_demo_2_2}
    \includegraphics[width=\columnwidth]{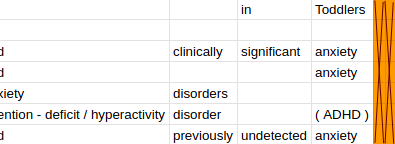}
\end{center}

\subsubsection{Shifting text}
\label{sec:usagescenario_interactive_shiftcell}

\begin{figure*}
    \centering
    \includegraphics[width=2\columnwidth]{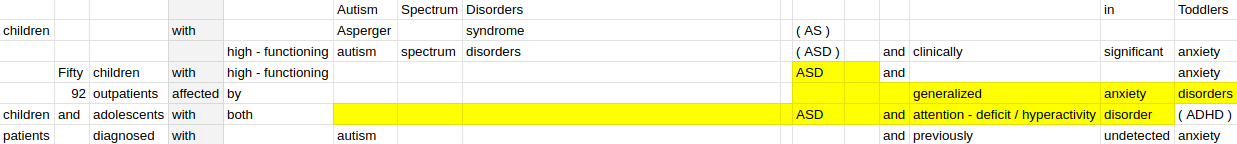}
    \caption{
        Usage Scenario: the full revised alignment generated by \sys\ based off of previous manual changes that Kim made.
        Cells that were changed by the system are highlighted in yellow.
    }
    \label{fig:usage_demo_3_0}
\end{figure*}

After a little bit of time exploring and modifying the alignment, Kim realizes that the texts can roughly be split into some description of the population demographics themselves, followed by some synonym of "diagnosed with", followed by descriptions of which conditions were being examined specifically.
Kim could now use this structure to do a finer breakdown of past participant groups.
For instance, they can extract a rough distribution of past study population demographics by age and group size by looking at the pre-"diagnosed with" sections of text, and correlate that with what conditions were studied by looking at the post-"diagnosed with" texts.

Even though Kim sees the big picture better now, they want to help their collaborators to see it even more easily.
Although the current alignment mostly reflects the main pattern already, there are still some mis-alignments so they manually shift some cells around to create separate "sections" in the alignment for each text component.
Specifically, Kim notes that there are two cells that both have the text "high - functioning" in them, and are both modifying the same type of noun.
Furthermore, one of these is in a column that otherwise entirely consists of cells containing the "diagnosed with" synonyms.
Kim decides to shift the "high - functioning" cells into the same column so they are aligned together and separated from the prepositions.

\begin{center}
    \label{fig:usage_demo_2_3}
    \includegraphics[width=\columnwidth]{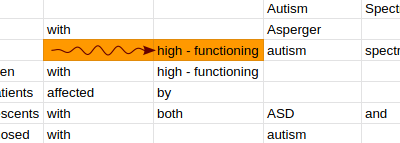}
\end{center}

\subsubsection{Locking columns}
\label{sec:usagescenario_interactive_lockcol}

After shifting the "high-functioning" cells to clarify the structure of their texts in the alignment, Kim wants \sys\ to recognize and enforce this structure.
They lock the columns that separate each of these sections in order to prevent the system from making any changes that contradict this structure.

\begin{center}
    \label{fig:usage_demo_2_4}
    \includegraphics[width=\columnwidth]{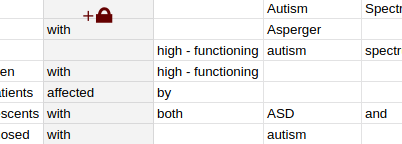}
\end{center}

\subsubsection{Automatic re-alignment}
\label{sec:usagescenario_realign}

After making manual adjustments to the cells, columns, and constraints on alignment structure, Kim clicks the "re-align" button to tell \sys\ to refine the alignment based on its current state and the column locking constraints they've specified.
\sys\ takes the current state of the alignment and tries to refine it more while staying within the constraints that Kim has given it (the column locking).
The alignment that \sys\ returns is something that Kim could take and share with their collaborators, or continue to interactively work with \sys\ to improve.

The system-revised alignment is viewable in Figure \ref{fig:usage_demo_3_0}.

\subsubsection{Inserting columns}
\label{sec:usagescenario_interactive_insertcol}

Kim thinks this alignment is mostly good enough to use as a way to organize the future texts they find.
However, they would like it to be easier to read so that they can share it with collaborators.
They notice how the study group sizes are aligned together with the group demographics and general descriptors.
To separate these for clarity, Kim inserts a new empty column to the left of the entire alignment and shifts the information about study group sizes into that column.

\begin{center}
    \label{fig:usage_demo_4_0}
    \includegraphics[width=\columnwidth]{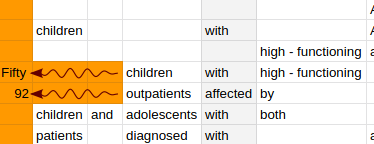}
\end{center}

At this point, Kim is satisfied with the state of the alignment.
If Kim changes their mind, it's possible to save the state of this alignment and load it back into \sys\ in the future to make more changes such as text shifting, merging columns together, or more automatic re-alignment.

\section{System Design}
\label{sec:systemdesign}

At a high level, \sys\ consists of three main components: an alignment algorithm which takes a collection of parallel texts as input and produces an "initial" alignment table, a stochastic hill-climbing search algorithm centered on an alignment quality heuristic which searches through the space of potential alignments given the user's specified changes and/or constraints, and the interface which allows the user to view and manually modify the text alignment and any alignment constraints.

\subsection{Initial Alignment Process}
\label{sec:systemdesign_initial}

The initial alignment algorithm functions to produce a basic alignment as a starting point to work with.
It isn't intended to produce the best-quality alignment.
Rather, the goal is to produce an initial alignment that is sufficiently reasonable for the state space search algorithm to improve on it quickly.

Given N input parallel texts, the initial alignment algorithm tokenizes each input text and generates a multiple alignment with N rows, where each row contains one of the input texts, readable from left to right.

It does this by repeatedly using the Smith-Waterman local alignment algorithm \cite{Smith1981IdentificationOC}.
Each run of the algorithm takes two input alignments (each of which contains one or more rows) and produces a single output alignment.
To generate a multiple alignment containing all N input texts, we initialize the final product by running Smith-Waterman local alignment on the first two input texts, then repeatedly aligning that result with the next input text until all N input texts are contained in the final alignment.
The input texts are added into the alignment in the same order as they are listed.

The Smith-Waterman algorithm itself uses "gap penalty" and "substitution matrix" heuristics to determine whether a given pair of columns from each input alignment should be aligned together or a gap should be inserted instead.
The gap penalty is calculated based on the length of the potential gap and penalizes multiple short gaps more than fewer long gaps.
The substitution matrix is calculated for each pair of columns that could be aligned together in the two inputs to Smith-Waterman algorithm, and scores how reasonable it would be to align the pair of columns together based on the L2 norm between the average phrase vector embeddings of the texts in each input column being considered for alignment together.
See appendix section "\nameref{sec:appendix_smithwaterman}" for exact details on how these are calculated for Smith-Waterman alignment in \sys.

For example, if given the following three lines of input in order:

\begin{enumerate}[nosep]
    \item \texttt{23 diabetics with flu}
    \item \texttt{six diabetic patients}
    \item \texttt{patients with flu}
\end{enumerate}

Smith-Waterman would take the first two input texts and produce the following:

\begin{equation}
    \label{eq:sysinitialalign1}
    \begin{array}{cccc}
    \texttt{23} & \texttt{diabetics} & \texttt{with} & \texttt{flu} \\
    \texttt{six} & \texttt{diabetic} & \texttt{patients}
    \end{array}
\end{equation}

Then it would add the third input text by aligning it to the existing partial alignment:

\begin{equation}
    \label{eq:sysinitialalign2}
    \begin{array}{cccc}
    \texttt{23} & \texttt{diabetics} & \texttt{with} & \texttt{flu} \\
    \texttt{six} & \texttt{diabetic} & \texttt{patients} \\
    & \texttt{patients} & \texttt{with} & \texttt{flu}
    \end{array}
\end{equation}

Note that this alignment is not an ideal alignment and that the order of the input texts can significantly affect the alignment that gets produced.

\subsection{Alignment State Space Search}
\label{sec:systemdesign_statesearch}

The basic initial alignment is a good starting point, but it often isn't an ideal alignment for two main reasons.
First (and most importantly), \sys\ is a mixed-initiative system and we want to accept and integrate user input, so we need to be able to take a given alignment that the user may have edited as input and improve it with a set of user-specified constraints.
Second, because the Smith-Waterman local alignment algorithm substitution matrix is calculated based on pairwise column comparisons, it is unable to use anything about the alignment that involves interactions between multiple columns to judge alignment quality.

To do this, we run a stochastic hill-climbing algorithm over the state space of all valid alignments (any alignment of the given input texts that obey all user-specified constraints) using an overall alignment quality heuristic.
The nodes in the search space graph are different potential alignments of the same input text and the edges are alignment modification operators that transform one alignment to a different one.

\subsubsection{Alignment modification operators}
\label{sec:systemdesign_statesearch_operators}

There are seven high-level types of operators: cell shift, column insert, column delete, column merge, cell merge, single-token column split, and trie column split.
Cell shift is the only operator used in the alignment state space search, as it combines well with the alignment quality heuristic.
The other four operations are available to the user, who can use them to manually edit an alignment on their side.

The sole alignment modification operator used in alignment state space search is the shift operator, which takes an input alignment, one or more cells of the alignment within the same column as each other, a direction to shift the cells in, and a distance to shift the cells by, and produces an output alignment where the cells are shifted to the left or the right by the given distance.
If shifting the cells as specified would move them over or through other cells that contain tokens, then it shifts all the cells that need to be "pushed" over together with the ones that were specified.

For example, given the initial alignment state in (\ref{eq:sysinitialalign2}), shifting the cell in column 2, row 3 to the right by 1 would produce:

\begin{equation}
    \label{eq:sysopshift1}
    \begin{array}{ccccc}
    \texttt{23} & \texttt{diabetics} & \texttt{with} & \texttt{flu} & \\
    \texttt{six} & \texttt{diabetic} & \texttt{patients} & \\
    & & \texttt{patients} & \texttt{with} & \texttt{flu}
    \end{array}
\end{equation}

Performing an additional shift of the cell in column 3, row 1 to the right by 1 would produce:

\begin{equation}
    \label{eq:sysopshift2}
    \begin{array}{ccccc}
    \texttt{23} & \texttt{diabetics} & & \texttt{with} & \texttt{flu} \\
    \texttt{six} & \texttt{diabetic} & \texttt{patients} & \\
    & & \texttt{patients} & \texttt{with} & \texttt{flu}
    \end{array}
\end{equation}

Note that in this example, the first shift operation arguably made the alignment worse.
Both operations together are needed to improve the alignment, performing only one of these operations by itself makes the alignment worse.
This is okay because we are doing a stochastic hill-climbing search, so the search algorithm will explore sub-par operations such as the first shift here.

An example of shifting multiple cells at once is if we start from the initial alignment state in (\ref{eq:sysinitialalign2}) and shift the cells in column 3, rows 1 and 3 to the right by 1 to produce:

\begin{equation}
    \label{eq:sysopshift3}
    \begin{array}{ccccc}
    \texttt{23} & \texttt{diabetics} & & \texttt{with} & \texttt{flu} \\
    \texttt{six} & \texttt{diabetic} & \texttt{patients} & \\
    & \texttt{patients} & & \texttt{with} & \texttt{flu}
    \end{array}
\end{equation}

Note that in this example, we can produce an alignment that's arguably better than the initial state using only a single operation, but is still slightly worse than alignment (\ref{eq:sysopshift2}) produced with a combination of two shift operations.
It would take one additional shift operation to transform alignment (\ref{eq:sysopshift3}) into (\ref{eq:sysopshift2}).

There are also six other alignment modification operators:column insert, column delete, column merge, cell merge, single-token column split, and trie column split.
These ended up combining poorly with the heuristic (unfortunately, the alignment quality heuristic still doesn't handle changes in alignment cell contents and column count very well) and aren't used in the system alignment search algorithm.
However, these operations are available for the user to use when manually editing an alignment.

Column insert takes an input alignment and a specified column of the alignment.
It produces an output alignment where there is one new empty column to the right of the specified column.

Column insert takes an input alignment and a specified column of the alignment.
It produces an output alignment where the specified column is entirely removed from the alignment.
If the specified column contains any text, the operation is invalid.

Column merge takes an input alignment and a specified column of the alignment.
It produces an output alignment where the specified column is combined with the one immediately to its right.
If the original column is already the rightmost column, then the operation is invalid.

For example, given the alignment states in both (\ref{eq:sysopshift2}) and (\ref{eq:sysopshift3}), performing a column merge on column 2 (merging columns 2 and 3) would produce:

\begin{equation}
    \label{eq:sysopcolmerge}
    \begin{array}{cccc}
    \texttt{23} & \texttt{diabetics} & \texttt{with} & \texttt{flu} \\
    \texttt{six} & \texttt{diabetic patients} & \\
    & \texttt{patients} & \texttt{with} & \texttt{flu}
    \end{array}
\end{equation}

In this example, this merge might be helpful because it combines the multi-token phrase "diabetic patients" in row 2 into one cell and aligns it with a synonymous phrase "diabetics" in row 1.

Cell merge takes an input alignment, a specified cell of the alignment, and a direction to merge it in.
It produces an output alignment where the specified cell is combined with the one immediately next to it, in the given direction.

For example, given the alignment state in (\ref{eq:sysopcolmerge}), performing a cell merge on the cell in column 3, row 1 to the right would produce:

\begin{equation}
    \label{eq:sysopcellmerge}
    \begin{array}{cccc}
    \texttt{23} & \texttt{diabetics} & & \texttt{with flu} \\
    \texttt{six} &  \texttt{diabetic patients} & \\
    & \texttt{patients} & \texttt{with} & \texttt{flu}
    \end{array}
\end{equation}

In this example, this merge is unhelpful because it changes a pair of perfectly aligned "with flu" texts into an imperfect alignment.
Furthermore, this change doesn't help any other section of texts.
However, notice how this cell merge can do something otherwise impossible without doing a combination of column insert, cell shift, and column delete operations.

Single-token column split takes an input alignment, a specific column of the alignment, and a flag for whether it should split from the left or the right.
It produces an output alignment where the specified column is separated into two columns.
One column contains either the leftmost or the rightmost token of each text in the original column as specified and the other column contains the rest of the original texts.
If the original column contains no texts with more than one token, then the operation is invalid.

For example, given the alignment state in (\ref{eq:sysopcolmerge}), performing a single-token column split on column 2 from the left would produce:

\begin{equation}
    \label{eq:sysopsinglesplitleft}
    \begin{array}{ccccc}
    \texttt{23} & \texttt{diabetics} & & \texttt{with} & \texttt{flu} \\
    \texttt{six} & \texttt{diabetic} & \texttt{patients} & \\
    & \texttt{patients} & & \texttt{with} & \texttt{flu}
    \end{array}
\end{equation}

And performing a single-token column split on column 2 from the right on the same input would instead produce:

\begin{equation}
    \label{eq:sysopsinglesplitright}
    \begin{array}{ccccc}
    \texttt{23} & & \texttt{diabetics} & \texttt{with} & \texttt{flu} \\
    \texttt{six} & \texttt{diabetic} & \texttt{patients} & \\
    & & \texttt{patients} & \texttt{with} & \texttt{flu}
    \end{array}
\end{equation}

In this pair of examples, alignment (\ref{eq:sysopsinglesplitright}) could be preferred over alignment (\ref{eq:sysopsinglesplitleft}) because alignment (\ref{eq:sysopsinglesplitright}) aligns all the main nouns together in column 3.
However, if we compare alignment (\ref{eq:sysopsinglesplitright}) with alignment (\ref{eq:sysopcolmerge}) from the column merge operator example, which alignment is "better" would depend much more on the user's individual needs and preferences.

Trie column split takes an input alignment, a specific column of the alignment, and a flag for whether it should split from the left or the right.
It builds a word trie starting from either the right or the left as specified containing all of the texts within the specified column and compresses the word trie into a phrase trie.
The final output is an alignment where the specified column is separated into two columns.
One column contains tokens in the first level of the phrase trie and the other column contains the rest of the original texts.
If the original column contains no texts with more than one token, then the operation is invalid.

For example, if we have this new example input alignment with a single column:

\begin{equation}
    \label{eq:sysoptriesplitsrc}
    \begin{array}{c}
    \texttt{2 young cancer patients} \\
    \texttt{15 adult cancer patients} \\
    \texttt{16 adult cancer patients} \\
    \texttt{2 young participants}
    \end{array}
\end{equation}


The process for performing a trie column split on this column from the right would start by building a word trie out of the column contents:

\begin{center}
    \label{fig:sysoptriesplitprocessfull}
    \begin{forest}
        for tree={
            circle,
            black,
            draw,
            fill=blue!40,
        }
        [{}
            [{}, edge label={node [midway, left] {\texttt{patients}}}
                [{}, edge label={node [midway, left] {\texttt{cancer}}}
                    [{}, edge label={node [midway, left] {\texttt{young}}}
                        [{}, edge label={node [midway, left] {\texttt{2}}}, label=below:{row1}]
                    ]
                    [,phantom]
                    [,phantom]
                    [{}, edge label={node [midway, right] {\texttt{adult}}}
                        [{}, edge label={node [midway, left] {\texttt{15}}}, label=below:{row2}]
                        [,phantom]
                        [{}, edge label={node [midway, right] {\texttt{16}}}, label=below:{row3}]
                    ]
                    [,phantom]
                ]
                [,phantom]
            ]
            [{}, edge label={node [midway, right] {\texttt{participants}}}
              [,phantom]
              [,phantom]
              [{}, edge label={node [midway, right] {\texttt{young}}}
                [{}, edge label={node [midway, right] {\texttt{2}}}, label=below:{row4}]
              ]
            ]
        ]
    \end{forest}
\end{center}

Next, compressing this into a phrase trie would produce:

\begin{center}
    \label{fig:sysoptriesplitprocesscompressed}
    \begin{forest}
        for tree={
            circle,
            black,
            draw,
            fill=blue!40,
        }
        [{}
            [{}, edge label={node [midway, left] {\texttt{cancer patients}}}
                [{}, edge label={node [midway, left] {\texttt{2 young}}}, label=below:{row1}]
                [,phantom]
                [,phantom]
                [{}, edge label={node [midway, right] {\texttt{adult}}}
                    [{}, edge label={node [midway, left] {\texttt{15}}}, label=below:{row2}]
                    [,phantom]
                    [{}, edge label={node [midway, right] {\texttt{16}}}, label=below:{row3}]
                ]
                [,phantom]
                [,phantom]
            ]
            [,phantom]
            [,phantom]
            [,phantom]
            [{}, edge label={node [midway, right] {\texttt{2 young participants}}}, label=below:{row4}]
        ]
    \end{forest}
\end{center}

Thus, the final output of performing a trie column split from the right on the one column of this input alignment would be:

\begin{equation}
    \label{eq:sysoptriesplitright}
    \begin{array}{cc}
    \texttt{2 young} & \texttt{cancer patients} \\
    \texttt{15 adult} & \texttt{cancer patients} \\
    \texttt{16 adult} & \texttt{cancer patients} \\
    & \texttt{2 young participants}
    \end{array}
\end{equation}

And performing a trie column split on the same input column from the left instead would produce this output alignment:

\begin{equation}
    \label{eq:sysoptriesplitleft}
    \begin{array}{cc}
    \texttt{2 young} & \texttt{cancer patients} \\
    \texttt{15 adult cancer patients} & \\
    \texttt{16 adult cancer patients} & \\
    \texttt{2 young} & \texttt{participants}
    \end{array}
\end{equation}

In these examples, neither alignment (\ref{eq:sysoptriesplitright}) nor alignment (\ref{eq:sysoptriesplitleft}) are good alignments.
Both alignments would require multiple additional operations to separate out the participant count, age group, and main noun into separate columns.
However, alignment (\ref{eq:sysoptriesplitright}) successfully separates the noun phrase "cancer patients" in rows 1, 2, and 3 from the participant count and age group details.
Alignment (\ref{eq:sysoptriesplitleft}) does not do this and is also harder to read at first glance in an alignment table format.

\subsubsection{Alignment search algorithm}
\label{sec:systemdesign_statesearch_algorithm}

We run a stochastic hill-climbing algorithm without random restarts on the state space of all possible alignments, where the edges of the search graph are the alignment modification operators that transform one potential alignment to another, and where the starting state is the current state of the alignment.

At each step in the stochastic hill-climbing "search" algorithm, we generate the full set of valid alignment modification operators based on the current state of the alignment (and also include "do nothing").
We remove all operators that conflict with any constraints on the alignment that the user has specified (i.e. if an operation would modify any locked columns).
Next, for each candidate operation, we compute the alignment that would result from applying it to the current alignment state and the alignment quality heuristic for that resulting alignment.
Finally, we select one of the candidate operations and apply it.
This repeats until the system has detected no operation would improve the heuristic for the last 2 iterations or until we have taken some specified number of steps, whichever happens first.
By default, we run a search after the initial alignment that caps at 50 steps, but the user is able to manually start a search from the current alignment state with a range of different step limits.

We select a candidate operation from the set of all valid operations by first deciding whether to take a random step or a greedy step.
By default, the probability of taking a greedy step is $0.5$ and probability of taking a random step is $0.5$.
If we take a greedy step, then we apply the operation that produces the highest heuristic score out of the entire set of valid operations.
If we take a random step, then we apply a random operation out of the entire set.

For example, if the current alignment state is:

\begin{equation}
    \label{eq:syssearchcurrent}
    \begin{array}{ccccc}
    \texttt{23} & \texttt{diabetics} & \texttt{with} & & \texttt{flu} \\
    \texttt{six} & \texttt{diabetic} & \texttt{patients} & \\
    & & \texttt{patients} & \texttt{with} & \texttt{flu}
    \end{array}
\end{equation}

The algorithm would start by generating every possible shift operation for the alignment.
(Note that we only consider the shift operation because it is the only operation used in the automatic alignment search. The split and merge operations are available for the user to manually edit an alignment but not used in the search algorithm.)
For example, these operations include (but are definitely not limited to):

\begin{enumerate}[nosep]
    \item Shift column 3, rows 2 and 3 to the right by 1
    \item Shift column 3, row 2 to the right by 1
    \item Shift column 3, row 1 to the right by 1
    \item Shift column 2, row 3 to the left by 1
    \item Shift column 4, row 3 to the left by 2
\end{enumerate}

If column 5 is locked, then we start by removing the first example operation ("Shift column 3, rows 2 and 3 to the right by 1") from the set of operations being considered because it conflicts with the lock constraint: it modifies the contents of column 5 and makes changes to the alignment that cross the "boundary" of that column lock.
Next, we calculate the results of applying each remaining modification operator and calculate the heuristic for each unique alignment that is produced.
If an operation doesn't actually modify the alignment (e.g. the fourth example operation), we discard it.
After that, we check if the alignment has hit the cutoff for deciding when to stop hill-climbing (if the best operation for the last N steps has been to "do nothing", or we have hit the step limit).
If we have not hit the step limit, we randomly decide whether to take a greedy step or random step.
In this example, if we take a greedy step, we would apply the third example operation because it produces the alignment with the highest heuristic score out of all operations tried.
If we take a random step in this example, we apply a random operation out of all the valid alignment modification operations.

\subsubsection{Alignment quality heuristic}
\label{sec:systemdesign_statesearch_heuristic}

The overall alignment quality heuristic used in the state space search algorithm is a weighted sum of three different score components.

When designing components to include in the heuristic, we roughly categorized each component as capturing one of two desired traits in an alignment: "conciseness", which penalizes alignments that are too large (e.g. alignments with many columns where there is only one token-containing row), and "coherence", which penalizes alignments with columns containing content that varies dramatically in meaning or purpose between different rows.

For example, here is an example alignment which is concise but not coherent:

\begin{equation}
    \label{eq:sysheuristicconcise}
    \begin{array}{cccc}
    \texttt{20} & \texttt{children} & \texttt{with} & \texttt{COVID} \\
    \texttt{five} & \texttt{male} & \texttt{adults} &
    \end{array}
\end{equation}

In this example alignment (\ref{eq:sysheuristicconcise}), the alignment is the smallest possible alignment to make if we limit each cell to one token maximum.
However, while column 1 successfully captures the patient count, column 2 aligns a patient age group ("children") with patient sex ("male"), while column 3 aligns a different patient age group ("adults") with a preposition.
Note that column 4 is perfectly coherent because there is only one row with text in that column, so it's impossible to have variation between texts in different rows.

Here is an example alignment containing the same data which is coherent but not concise:

\begin{equation}
    \label{eq:sysheuristiccoherent}
    \begin{array}{ccccccc}
    \texttt{20} & & & \texttt{children} & & \texttt{with} & \texttt{COVID} \\
    & \texttt{five} & \texttt{male} & & \texttt{adults} & &
    \end{array}
\end{equation}

While each column in alignment (\ref{eq:sysheuristiccoherent}) has zero variation between the texts that they contain, that's only because each column only contains one text.
This alignment has too many gaps and needlessly separates tokens that are thematically similar (e.g. "children" and "adults") into separate columns.

And finally, here is an example alignment of the same data that ideally combines both conciseness and coherence:

\begin{equation}
    \label{eq:sysheuristicboth}
    \begin{array}{ccccc}
    \texttt{20} & & \texttt{children} & \texttt{with} & \texttt{COVID} \\
    \texttt{five} & \texttt{male} & \texttt{adults} & &
    \end{array}
\end{equation}

There are three main score components that contribute to the overall quality heuristic, two of which encourage the alignment to be concise and one which encourages the alignment to be coherent:

\begin{itemize}[nosep]
    \item
        \textit{Number of total columns} (conciseness).
        
        This is calculated by counting how many columns the alignment contains, then dividing it by the minimum possible alignment column count (the number of columns that the longest single row of the alignment would occupy).
        This normalizes it so that the subscore is comparable between alignments with modified tokenization.
    
    \item
        \textit{Number of filled columns} (conciseness).
        
        This is the same as the number of columns, except instead of counting the total number it only counts columns that contain any text in them at all.
        This penalizes alignments with empty columns.
    
    \item
        \textit{Embed variance / Similarity of texts within each column} (coherence).
        
        For each of the columns in the alignment, we use FastText \cite{Mikolov2018AdvancesIP, Bojanowski2017EnrichingWV, Joulin2017BagOT} to find the set of word embeddings (or phrase embeddings, calculated as the average of each of the word embeddings in a phrase) and calculate the variance of each embedding from the column average.
        We also calculate the "relevance" of each column as the fraction of rows represented by any tokens within each column.
        The overall "embed variance" component score is the dot product of the per-column word embedding variance and the "relevance" score weights.
\end{itemize}

For example, if we have this alignment (note that in this example, unlike previously, we use "$\square$" to represent an empty cell for readability):

\begin{equation}
    \label{eq:sysheuristiccomponentsrc}
    \begin{array}{ccccc}
    \texttt{23} & \texttt{diabetics} & \texttt{with} & \square & \texttt{flu infection} \\
    \texttt{six} & \texttt{diabetic patients} & \square & \square & \square \\
    \square & \texttt{patients} & \texttt{with} & \square & \texttt{flu}
    \end{array}
\end{equation}

The "number of total columns" component would be calculated:

\begin{equation}
    \frac{\text{total column count}}{\text{cells in longest row}} = \frac{5}{4} = 1.25
\end{equation}

The "number of filled columns" component would be calculated:

\begin{equation}
    \frac{\text{filled column count}}{\text{cells in longest row}} = \frac{4}{4} = 1
\end{equation}

The "embed variance" component would be calculated:

\begin{equation}
    \begin{pmatrix}
        \text{col1 relevance} \\
        \text{col2 relevance} \\
        \text{col3 relevance} \\
        \text{col4 relevance} \\
        \text{col5 relevance}
    \end{pmatrix}
    \cdot
    \begin{pmatrix}
        \text{col1 variance} \\
        \text{col2 variance} \\
        \text{col3 variance} \\
        \text{col4 variance} \\
        \text{col5 variance}
    \end{pmatrix}
\end{equation}

\begin{equation}
    = 
    \begin{pmatrix}
        2/3 \\
        3/3 \\
        2/3 \\
        0/3 \\
        2/3
    \end{pmatrix}
    \cdot
    \begin{pmatrix}
        \text{Tr}(\text{Cov}(e_{col1})) \\
        \text{Tr}(\text{Cov}(e_{col2})) \\
        \text{Tr}(\text{Cov}(e_{col3})) \\
        \text{Tr}(\text{Cov}(e_{col4})) \\
        \text{Tr}(\text{Cov}(e_{col5})) \\
    \end{pmatrix}
\end{equation}

where $\text{Tr}(y)$ calculates the trace of an input matrix $y$, $\text{Cov}(x)$ calculates the covariance matrix of an input matrix $x$, and $e_{coln}$ is a matrix where each row contains the word (or averaged phrase) embedding vector for a token-containing row in column $n$ of the alignment.

The final quality heuristic is a combination of these score components.

For exact details on how these score components are combined, see appendix section "\nameref{sec:appendix_heuristic}".

We will now describe how we selected the heuristic.

In the process of refining a heuristic, we considered a wide variety of alignment features to use as components.
(See appendix section "\nameref{sec:appendix_heuristiccandidates}" for details on which features were considered and how each of them were calculated.)

We also built a test set\footnote{\href{https://github.com/cephcyn/alignpaper/tree/master/testcases}{https://github.com/cephcyn/alignpaper/tree/master/testcases}} containing pairs of comparable alignments to evaluate each component and candidate heuristic weighting.
We used the EBM-NLP dataset \cite{Nye2018ACW} containing annotated "Patient" text spans from medical abstracts as a source for parallel texts.

First, we created several groups of similarly themed texts (e.g. all texts mentioning cancer, or all texts mentioning studying cats or dogs).

Second, we built several alignments for each of these groups.
This included hand-made "ideal" alignments, as well as system-generated alignments from Smith-Waterman multiple alignment.
This also included alignments containing only small subsets of entire text groups, together with full alignments.

Third, we manually created one or more modified alignments based on each of these alignments to either (subjectively) improve them or worsen them.
These modifications ranged from simple test cases (e.g. a single shift to emphasize that words with similar meanings or parts of speech should be aligned together, or merging two columns with synonymous phrases of different token lengths together) to "holistic" improvements which would require multiple alignment modification operators to step from one to the other.
This created many pairs of "better" and "worse" alignments that we used as individual pairwise comparison test cases for the overall heuristic.
A good heuristic would score the "better" alignment in each pair higher than the "worse" alignment more often than a bad heuristic.

Using our list of initial candidate components and the pairwise comparison test set, we started with an initial candidate heuristic and manually varied the weighting of each component in the heuristic until we were unable to improve how well it performed on the test set.

\subsection{User Interface}
\label{sec:systemdesign_interface}


Users have a range of options for interacting with the alignment system, including providing input parallel texts for the system to align, making manual changes to the current alignment state, modifying alignment search constraints, and instructing the system to perform an alignment optimization search based on the edits and constraints they have specified.

\begin{figure}[h!]
    \centering
    \includegraphics[width=1\columnwidth]{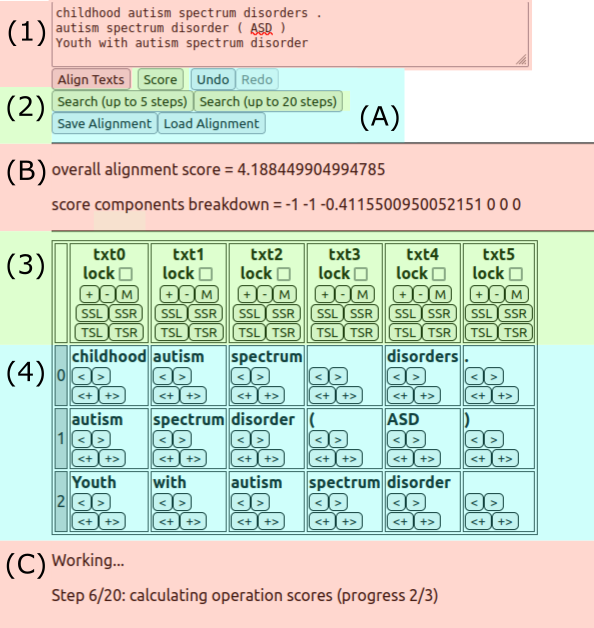}
    \caption{
        User Interface: an example of the general user interface in \sys.
        Each highlighted and annotated section in this figure contains a group of interface controls or output areas, which are described in more detail separately.
    }
    \label{fig:userinterface_basic}
\end{figure}

We now describe the contents of the main interface, as shown in Figure \ref{fig:userinterface_basic}.

Section (1) takes user input to create an alignment from scratch.
Users can type or copy-paste input into the text field.
Each line of the input is considered a separate phrase to be aligned, and will become its own row.

Section (2) contains controls to score the alignment, as well as to run re-alignment to improve the current alignment.
The two different re-alignment buttons have different numbers of maximum state space search steps: the fewer steps, the less time the search algorithm will spend exploring the state space.

Sections (3) and (4) are part of the alignment visualization itself, with different controls in different sections.

Section (3) contains manual alignment controls pertaining to individual columns.
Each column contains a "lock" checkbox which indicates whether the column should remain unchanged during alignment search.
Each column also contains multiple manual modification buttons.
"+" inserts a new empty column to the right of the indicated column.
"-" deletes the indicated column.
"M" merges the indicated column with its neighbor on the right.
"SSL" and "SSR" perform single-token column split from the left and the right, respectively.
"TSL" and "TSR" perform trie column split from the left and the right, respectively.

Section (4) contains manual alignment controls for individual cells.
"<" and ">" shift the indicated cell to the left and the right, respectively.
"<+" and "+>" merge the indicated cell to the left and the right, respectively.
If the cell that it attempts to merge with is empty, then clicking "<+" or "+>" is equivalent to clicking "<" or ">".

Section (A) allows users to load/save the state of an alignment, or undo/redo the changes they have made to an alignment.

Section (B) shows the heuristic score for the current alignment state.

Section (C) shows the current running status of \sys.

The user can also specify constraints for the alignment search algorithm to take into consideration.
They can lock a column of the alignment, which prevents the alignment search algorithm from performing any operations that would change any cell within that column, as well as from changing the horizontal relationship between cells within that column and any adjacent columns.

\begin{figure}[h!]
    \centering
    \includegraphics[width=0.6\columnwidth]{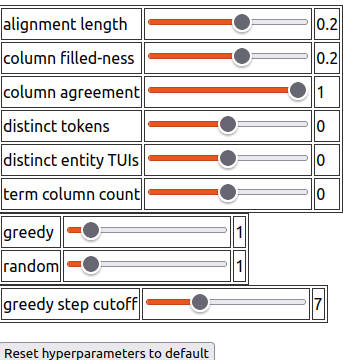}
    \caption{
        User Interface: the controls for adjusting hyperparameters.
    }
    \label{fig:userinterface_poweruser}
\end{figure}

Furthermore, users can experiment with the system algorithm internals themselves.
While \sys\ has default values for hyperparameters (e.g. greedy step cutoff and greedy vs. random probability in the stochastic hill-climbing algorithm, or the weighting for each score component in the alignment heuristic), a power user can change these hyperparameters to explore system functionality or to guide the system towards different alignment styles.
The hyperparameter control interface is shown in Figure \ref{fig:userinterface_poweruser}.

\section{Conclusions}
\label{sec:discussion}

Sensemaking over collections of related and comparable texts is something that is important for scientists and science journalists to do, as part of their background reading and understanding.
This sensemaking is often time-consuming, as well as being very difficult to fully automate.
In this paper, we presented \sys, which combines an automated alignment and search algorithm that can help align comparable texts in a way that makes them easier to collectively summarize, with a mixed-initiative design that keeps a human researcher in the loop and folds their contextual knowledge into the process.
It's important to note that while we demonstrated using \sys\ with medical examples and PICO in this paper, \sys\ could be used in other domains as well, such as compiling the improvements to or applications of different machine learning models.

As always, there are a range of directions that this work could be expanded in.

One straightforward next step is to perform a user evaluation of how usable and helpful \sys\ is for sensemaking.
This would require picking a user and task domain, as well as figuring out how best to evaluate usability and prevent any biases coming from system slowness.

Another major potential area of improvement is the alignment quality heuristic itself, especially because it drives the core interaction loop in \sys.
This could be done by improving the heuristic selection process by building a more comprehensive and thorough test set, or experimenting with a wider range of alignment quality metrics.
Furthermore, we could attempt to improve the heuristic based on the user interactions themselves.
Individual "sessions" of user interaction with \sys\ could be used as some kind of training data to evaluate the quality of an alignment for future "sessions".

Additionally, we could expand what types of interaction the user is able to have with \sys.
For example, users may want to use an existing alignment as a base to align new comparable texts to.
We could also expand what types of alignment modifications or constraints a user is able to make, ideally in a way that improves usability.

Finally, it would be interesting to examine more closely how \sys\ and text alignment algorithms in general could use generated sentence structure trees as part of the process.
In English, while there are common ways of ordering sentence components such as subjects and objects in written text, these aren't strict requirements.
If a collection of input texts contain some with such varying structures, it may be impossible to align them all well without reordering any of the inputs.
This challenge might be even more difficult in other languages.
However, taking advantage of sentence structure trees to help with reordering input texts could preserve semantics while improving alignment potential.

\section{Acknowledgments}
\label{sec:acknowledgements}

Thanks to Jolie Zhou and Travis McGaha for helpful comments.

This material is based upon work supported by ONR grant {N00014-18-1-2193}, NSF RAPID grant 2040196, the WRF/Cable Professorship, and the Allen Institute for Artificial Intelligence (AI2).

%
%
%
%

\balance{}

\section{Appendix}
\label{sec:appendix}

\subsection{Smith-Waterman alignment heuristics}
\label{sec:appendix_smithwaterman}

The gap penalty score ($S_g$) is calculated based on potential gap length ($l$):

\begin{equation}
    S_g = -1 * (1 * \min(l, 1) + 0.1*\max(l-1, 0))
\end{equation}

The substitution matrix (actually more of a function, as there are an infinite number of potential inputs to be aligned together) calculates a score ($S_a$) for how "alignable" two columns from two different input alignments would be.
It is calculated using the two sets of texts, one set from each input alignment being combined together.
We denote these as $t_{a1} \dots t_{an}$ for the $n$ texts in the first input alignment $a$, and $t_{b1} \dots t_{bm}$ for the $m$ texts in the second input alignment $b$.

If both input sets contain at least one text that can be parsed into a vector embedding, then the substitution score calculation starts by calculating a normalized vector embedding for every phrase it is able to, such that we have $v_{a1} \dots v_{aq}$ for the $q$ texts in the first input alignment that can be parsed into embeddings, and $v_{b1} \dots v_{br}$ for the $r$ texts in the second input alignment that can be parsed into embeddings, and $|v - \vec{0}| = 1$.
Then $S_a$ is calculated:

\begin{equation}
    S_a = 10 * (6 - |
        \sum_{i=1}^{q} \frac{v_{ai}}{q} -
        \sum_{j=1}^{r} \frac{v_{bj}}{r}|)
\end{equation}

Sometimes, texts contain serious typos or simply don't make sense.
If neither input set contains any texts that can be parsed into a vector embedding (so if they are entirely unrecognized tokens) then $S_a$ is calculated using a normalized adaptation of Levenshtein distance averaged across all of the inputs:

\begin{equation}
    S_a = 60 * (1 - \frac{1}{nm}*
    \sum_{i=1}^{n} \sum_{j=1}^{m}
    \frac{
        \text{Levenshtein}(t_{ai}, t_{bj})
    }{
        \max(\text{length}(t_{ai}), \text{length}(t_{bj}))
    })
\end{equation}

\subsection{Overall alignment quality heuristic}
\label{sec:appendix_heuristic}

The overall alignment quality heuristic ($S_q$) is calculated as:

\begin{equation}
    S_{q} =
    \begin{pmatrix}
        -1 * w_{col} \\
        -1 * w_{fcol} \\
        -1 * w_{embed} \\
        w_{bias}
    \end{pmatrix}
    \cdot
    \begin{pmatrix}
        s_{col} \\
        s_{fcol} \\
        {s_{embed}}^2 \\
        1
    \end{pmatrix}
\end{equation}

where:

\begin{itemize}[nosep]
    \item 
        $w_{col}$ defaults to 0.2.
        
        This is negative because fewer columns is more concise.
    
    \item 
        $s_{col}$ is the "number of columns" score component.
    
    \item 
        $w_{fcol}$ defaults to 0.2.
        
        This is negative because fewer columns is more concise.
    
    \item 
        $s_{fcol}$ is the "number of filled columns" score component.
    
    \item 
        $w_{embed}$ defaults to 1.
        
        This is negative because lower embed variance means phrases are more similar.
    
    \item 
        $s_{embed}$ is the "similarity / embed variance" score component.
    
    \item 
        $w_{bias}$ defaults to 5.
        
        This bias "weight" exists because a positive score is easier to intuitively compare than a negative score.
\end{itemize}

Note that $w_{col}$, $w_{fcol}$, and $w_{embed}$ are adjustable in the user interface.

\subsection{Expanded list of alignment heuristic components considered}
\label{sec:appendix_heuristiccandidates}

In addition to the three heuristic score components described in the main body of this paper (number of total columns, number of filled columns, and embed variance), we considered and tested out multiple other component ideas.
Some of these are implemented in the interface, but temporarily disabled to save computation time.
Here, we describe the other score components tested.

\begin{itemize}[nosep]
    \item
        \textit{Number of distinct \_\_\_ per column} (coherence).
        
        This would be a count of how many unique text phrases, phrase parts-of-speech, tokens, token parts-of-speech, entities, or entity types are in each column.
        Calculations on unique phrases and tokens were based on built-in string computation, while varying ML models were used to predict parts-of-speech and tag entities.
        These all target the same basic idea as embed variance, with different emphasis on runtime and flexibility (e.g. entity prediction only works in a medical context).
        A major shared weakness is that each acknowledges only a binary sense of similarity.
    
    \item
        \textit{Number of rows represented per column} (conciseness).
        
        This would be calculating how "filled" each column was.
    
    \item
        \textit{Count of how much each term is spread across columns} (conciseness).
        
        This would calculate how many columns that a given token, phrase, or group of phrases was represented in.
        It would need to be combined with a list of these phrases to check for, as well as how much to weight each of these phrases' importance.
        This was intended to discourage significant terms from being spread across too many columns in the alignment.
\end{itemize}

\balance{}

\bibliographystyle{SIGCHI-Reference-Format}
\bibliography{references}

\end{document}